\documentclass[twocolumn,10.5pt]{article}

\usepackage{graphicx}
\usepackage{dcolumn}
\usepackage{bm}
\usepackage[utf8]{inputenc}
\usepackage[T1]{fontenc}
\usepackage{mathptmx}
\usepackage{authblk}
\usepackage{amsmath}
\usepackage{amssymb}
\usepackage{xcolor}
\definecolor{lightblue}{HTML}{0264ab}

\newcommand{\tprime}{t^\prime}
\begin{document}

\title{Memory Control of Ice Growth \\During Non-Equilibrium Freezing of Water}
% Force line breaks with \\
\author[1]{Abhigyan Hazarika}
\author[2]{Sudeep N. Punnathanam}
\author[3]{Biman Bagchi}
\author[4]{Prabal K. Maiti}
\affil[1]{Center for Condensed Matter Theory, Department of Physics, \\Indian Institute of Science, Bangalore, Karnataka, India}
\affil[2]{Department of Chemical Engineering, \\Indian Institute of Science, Bangalore, Karnataka, India}
\affil[3]{Solid State and Structural Chemistry Unit, \\Indian Institute of Science, Bangalore, Karnataka, India}
\affil[4]{Center for Condensed Matter Theory, Department of Physics, \\Indian Institute of Science, Bangalore, Karnataka, India\\\texttt{maiti@iisc.ac.in}}

\date{\today}
\maketitle
\begin{abstract}
Freezing of supercooled water is a classic non-equilibrium problem, yet the influence of thermal history on crystallization remains unclear. Using molecular dynamics simulations with the TIP4P/Ice model, we investigate how the initial temperature $T_i$ shapes freezing following rapid quenching to 250 K. By monitoring the evolution of hydrogen-bonded ring structures, we find a non-monotonic dependence of the freezing time $t_F$ on $T_i$, with the slowest crystallization occurring near 300 K. Remarkably, this means that initially hotter water can freeze faster than cooler water, a molecular-scale analogue of the Mpemba effect. A non-stationary generalized Langevin equation framework shows that two-time memory kernels retain information about the system’s thermal past, directly influencing crystallization dynamics. Structural analysis further reveals that five-membered rings act as kinetic traps, while correlations among ring types regulate the accessibility of ice-like motifs. These results uncover a molecular origin of memory-driven freezing and establish structural memory as a key driver of non-equilibrium phase transitions.
\end{abstract}

% SECTION: INTRODUCTION
\section{Introduction}
The process of freezing can be classified into two fundamental categories: equilibrium and non-equilibrium freezing. Equilibrium freezing describes phase transitions that occur near the freezing temperature, where the system remains close to thermodynamic equilibrium. In contrast, non-equilibrium freezing happens when a liquid is supercooled far below its freezing point, causing any subsequent solidification to be non-equilibrium in nature. While equilibrium freezing is a well-studied phenomenon, its non-equilibrium counterparts are less understood due to their theoretical and experimental complexity.

In classical molecular dynamics, simulating the equilibrium freezing of water with realistic potentials poses significant computational challenges due to the long timescales of spontaneous crystallization. A landmark study by Matsumoto et al \cite{Matsumoto2002} showed that observing freezing in atomistic simulations requires the formation of a stable nucleus, a rare event due to the vast number of configurations available in water's hydrogen-bond network. This makes spontaneous nucleation computationally prohibitive. Later, Johnston and Molinero \cite{Johnston2012} achieved crystallization with a coarse-grained mW model, but at cooling rates orders of magnitude faster than in experiments. Further studies revealed that realistic crystallization timescales are computationally inaccessible, creating a gap between simulation and reality \cite{Espinosa2016}. This limitation has spurred the development of enhanced sampling methods and coarse-grained models, though achieving homogeneous crystallization in equilibrium simulations remains a major hurdle.

Non-equilibrium freezing covers a wide range of behaviors. When liquid water is cooled rapidly, it can vitrify into amorphous ices that lack long-range order and exhibit a glass transition \cite{Binder2014,Bachler2021,Melillo2024}. These disordered states are kinetically stabilized. In nature and industry, non-equilibrium effects are harnessed for innovation. For instance, antifreeze proteins (AFPs) bind to ice crystals, depressing the freezing point and modifying crystal shape to prevent cellular damage \cite{Knight2009}. In geophysics, glacial ice undergoes regelation, i.e. cyclic melting and refreezing: which affects glacier movement \cite{Hubbard1993,Meyer2024}. Furthermore, non-equilibrium freezing on mineral surfaces and in supercooled atmospheric droplets plays a crucial role in climate science \cite{Sosso2016,Bogdan2018,Murray2012}. These examples show that non-equilibrium freezing is a widespread and significant physical process.

A compelling feature of non-equilibrium processes is the presence of memory effects, where a system's behaviour reflects its thermal and structural history. In glasses, the concept of a fictive temperature is used to describe how a system retains a signature of its cooling protocol \cite{Badrinarayanan2007,Mauro2009}. The fictive temperature represents the temperature at which the system's arrested structure would be in equilibrium, linking its microscopic state to its history. This \textit{ipso facto} "memory" is a common feature in systems with complex energy landscapes, where relaxation processes depend on the pathway taken.

Insights from general memory phenomena in complex systems further illuminate non-equilibrium processes. The Kovacs effect, first observed in polymer glasses, describes a non-monotonic return to equilibrium after a temperature change, reflecting the system's memory of its prior state \cite{Kovacs1979}. This effect has since been identified in diverse systems, including glass-forming liquids \cite{Mossa2004}, fragile glassy polymers \cite{Aquino2006}, the ferromagnetic Ising model \cite{RuizGarcia2014}, and driven granular gases \cite{Prados2014}. Another key example is the Mpemba effect, where a hotter system can freeze faster than a cooler one, originally observed in water experimentally \cite{mpemba1969,Burridge2020,Tang2022,Janni2026} and in numerical studies \cite{Kier2013,Zhang2014,Jin2015,Tao2016,Ghosh2025}. This counter-intuitive behavior is now recognized as a phenomenon of complex energy landscapes, appearing in systems such as the Ising model \cite{Lu2017,Vadakkayil2021}, spin systems \cite{Chatterjee2024}, granular fluids \cite{Lasanta2017}, colloidal systems \cite{Kumar2020,Biswas2023}, and trapped ion systems \cite{Zhang2025}. Of note is the one-dimensional Ising model, that exhibits both Kovacs and Mpemba effects based on the type of interaction being modelled. Considering ferromagnetic interactions leads to the exhibition of the Kovacs effect when the system undergoes a temperature cycle and the subsequent relaxation process \cite{RuizGarcia2014}. When antiferromagnetic interactions are considered with an external field, the model exhibits the Mpemba effect, where initially far-from-equilibrium states relax quicker than initially near-equilibrium states \cite{Lu2017}. The common thread between these two seemingly unrelated effects is that they feature non-trivial relaxation of energy to equilibrium. The ferromagnetic example displays the effect in a regime where non-exponential glassy relaxations come into play. In case of the antiferromagnetic example, the energy landscape of the model system was demonstrated to be rugged, consisting of many metastable wells, which slow down the eventual relaxation, a description also fitting for water, a complex liquid. The two-dimensional variant of the Ising model with ferromagnetic interactions was also shown to exhibit the Mpemba effect\cite{Vadakkayil2021,Chatterjee2024}, when paramagnetic initial configurations were cooled to ferromagnetic temperatures. Both the Kovacs and Mpemba effects underscore the importance of memory effects in out-of-equilibrium systems, where the past influences the present. 

The primary goal of this study is to investigate the kinetics of non-equilibrium freezing in supercooled liquid water. By extracting freezing timescales from numerous molecular dynamics (MD) trajectories with varied initial conditions, one can quantify crystallization kinetics and its dependence on history. At a large supercooling of  $\approx15$K or more, it was shown that interfacial crystal growth is diffusion-limited, meaning it is controlled by the rate of molecular rearrangements in the liquid \cite{Weiss2011}. It was also shown that the thermodynamic driving force for freezing grows with decreasing temperature, and the limiting factor comes from the molecular diffusivity, which decreases in an Arrhenius-like fashion that counteracts the increasing driving force. The MD simulations in this study, conducted at a supercooling of $\approx 20$K, operate within this regime, allowing one to computationally probe the interplay between thermal history, the energy landscape, and diffusion-limited growth in producing memory effects.

A timescale separation also distinguishes between nucleation timescales and the interfacial growth kinetics within the scope of this work. A study of the nucleation rates for the TIP4P/Ice water model \cite{tip4p-ice} carried out by Espinosa et al. \cite{Espinosa2016} showed that this model reproduces the nucleation rate (J) faithfully with experiment. However, the downside they showed is that the crystallization timescales associated with typical MD system sizes will be orders of magnitude higher than what we have examined here ($\sim 100$ns) with the direct-coexistence system.

Using the TIP4P/ice water model \cite{tip4p-ice}, the freezing of liquid water into hexagonal ice (ice Ih) was simulated. The time required for the system to freeze completely was measured for various initial liquid water temperatures to assess the kinetics. To observe this process in unbiased MD, the direct coexistence method was employed, where ice and liquid phases are kept in contact within the same simulation cell (Figure \ref{fig:figure1}A) \cite{Fernandez2006}. By running simulations in the NPT ensemble at $250 K$, one drives the system toward the ice phase in a non-equilibrium manner. The kinetics of this growth are shaped by multiple factors, particularly the microscopic structural arrangement at the initial temperature. Quenching from different temperatures traps different initial structures, placing the system on distinct regions of its energy landscape. This sensitivity of freezing time to initial temperature offers crucial insights into the kinetics of ice growth, analogous to the influence of complex energy landscapes seen in the Mpemba effect in other models \cite{Lu2017}.

Quantifying the influence of a system's history on its dynamics is a central challenge in non-equilibrium statistical mechanics. The non-stationary generalized Langevin equation (nsGLE) framework \cite{Netz2024,Ayaz2022} offers a powerful approach by extending the projection operator formalism to systems far from equilibrium \cite{Meyer2017,Glatzel2021}. The nsGLE replaces the time-invariant kernels of conventional GLEs with time-dependent memory kernels, capturing evolving, history-dependent effects. Meyer et al. established a link between these memory timescales and the duration of phase transformations \cite{Meyer2021}. Building on this, memory kernels were computed from the MD trajectories in this work. By comparing them to the observed kinetics, the strength of memory effects can be quantified and connected to the microscopic rearrangements of the hydrogen-bond network.

Finally, this study examines the underlying molecular-level mechanisms driving the freezing process, focusing on ring structures within the hydrogen-bond network (HBN) of liquid water. The importance of five-membered rings in water’s structure and dynamics has long been recognized. Speedy \cite{Speedy1984,Speedy1985} first proposed that they contribute to the metastability and anomalies of supercooled water, a view later substantiated by Belch, Sceats, and Rice \cite{Belch1987} through detailed analysis of the hydrogen-bond network. More recent studies have clarified their role in the high-density liquid (HDL)–low-density liquid (LDL) scenario: Russo and Tanaka \cite{Russo2014} identified five-membered rings as key to the two-state mixture description of water, while Martelli \cite{Martelli2019} showed that frustration between five and six-membered rings is maximized under supercooling, suppressing crystallization. Simulations by Bullock and Molinero showed that crystallization is preceded by the formation of low-density liquid (LDL) domains \cite{Bullock2013}, which have a density similar to ice Ih \cite{Sciortino2024}. These motifs thus serve as structural antagonists: five-membered rings stabilizing HDL-like arrangements and hindering the formation of six-membered, ice-like order associated with LDL-like domains. At temperatures below freezing, their persistence slows crystallization by acting as kinetic bottlenecks, while their weakening at higher initial temperatures allows faster growth. In this way, five-membered rings encode structural memory in the hydrogen-bond network, providing a molecular origin for history-dependent freezing kinetics and Mpemba-like behavior. 

Through a detailed analysis of the ring network's evolution in both equilibrium and non-equilibrium simulations, it will be demonstrated that the persistence and temporal cross-correlations of these ring motifs are linked to emergent memory effects in freezing kinetics. It will also be shown that the five-membered ring plays a key role due to its inherent structural stability, which makes it a kinetic bottleneck in the liquid-to-ice transformation.

\section{Methods}
%% \label{}
All water molecules were parametrized by the TIP4P/Ice model \cite{tip4p-ice}, a 4-point rigid water model. The corresponding force field parameters have been listed in Table S1 of the \textit{Supplementary Information} (SI). The melting point of ice for this water model (272K) is close to the experimental melting point (273K), making it a good candidate for simulating the freezing and formation of ice. The GROMACS 2024 \cite{GROMACS2015} molecular dynamics package was used to simulate all the systems in this study. An integration timestep of 2.0 fs was used, applying the leap-frog integration scheme. Atomic bonds were constrained using the LINCS algorithm \cite{LINCS}. Short-range non-bonded interactions were treated with a shifted potential function with a 1.0 nm cutoff distance, and long-range electrostatics were computed using Particle Mesh Ewald (PME) electrostatics, with a real space cutoff of 1.0 nm and a Fourier grid spacing of 0.12 nm. 

\subsection{Generation of the starting configuration}
The initial configuration of the hexagonal ice Ih comprising 432 water molecules was generated through GenIce \cite{Genice2}, a software package that algorithmically generates various ice structures. The generated structure was subjected to an energy minimization using steepest descent algorithm to resolve high energy contacts. In the next step, an NPT equilibration was performed at 1 atm pressure and 10 K temperature for a period of 1 ns. The initial velocities were sampled from a Boltzmann distribution corresponding to a temperature of 10 K. The pressure and temperature were controlled using a Berendsen barostat \cite{Berendsen1984} with a time constant of 1.0 ps and a stochastic rescaling thermostat \cite{bussi2007} with the same time constant respectively. The pressure coupling was chosen to be in the anisotropic mode. Finally, the system was heated to 250 K through a linear ramp with a heating rate of 0.1 K/ps followed by a further equilibration at 250 K for a period of 4 ns. The final dimensions of this ice box were 2.36 nm, 2.22 nm and 2.72 nm in the x, y and z directions respectively. 

The liquid configuration consisted of 2592 molecules of water placed in a simulation box having x and y dimensions equal to that of the ice configuration, and the box length in the z direction equal to 15.9 nm so as to maintain the liquid water density of 1.0 g/cm$^3$. The liquid phase was equilibrated by energy minimization followed by a 2 ns NVT equilibration run with the same thermostat, keeping the reference temperature at 300K. 

Once the ice and liquid configurations were obtained, they were brought in contact by connecting them side-by-side, keeping the interface perpendicular to the the z-axis. Thus, total number of water molecules in this system was $2592 + 432 = 3024$. Another steepest descent minimization was performed, after which the resulting structure was taken as the starting configuration, shown in Figure \ref{fig:figure1}A, for the freezing study. 

\subsection{Heating and quenching protocol}
To create the conditions to examine the dependence of initial temperature, the liquid water has to be heated to said temperatures, and cooled subsequently to a lower temperature to freeze it. In order to ensure that the ice phase remains frozen and does not melt under the high temperature coupling, the ice molecules were position-restrained through a harmonic energy penalty while the thermostat was set to the desired initial temperature ($T_i$, also referred to as \textit{starting temperature}) with a time constant of 1 ps. The Berendsen anisotropic barostat $(\tau = 1ps)$ was utilized, with pressure coupling applied only on the z-direction. After running for 2 ns, it is ascertained that the liquid phase has reached the required temperature by monitoring the energies of the system. At this point, the production run was initiated. The thermostat was turned down to 250 K (referred to as 'quenching' in this work), with time constant 0.1 ps, a choice derived from Jin and Goddard \cite{Jin2015}. Under these conditions, the system was simulated for another 180 ns. A visual representation of the time evolution of this system is shown in Figure \ref{fig:figure1}B. 

\subsection{Ring analysis of hydrogen-bond network}
The presence and behavior of hydrogen bonds are crucial to the various properties of water, influencing its various thermodynamic and structural properties \cite{Texeira1993}, \cite{Luzar1996}. In liquid water, the possibility of hydrogen bonding allows the molecules to form hydrogen bond networks of different morphologies, which give rise to the various anomalous properties of this liquid \cite{Stillinger1980}. A study of this network is essential to achieving a more granular understanding of not only the microscopic but also the macroscopic behaviour of water. To that end, a topological approach can be taken, wherein the hydrogen bond network of water can be defined as an undirected graph, treating atoms as vertices, and the hydrogen bonds as edges. With this formalism, graph algorithms can be applied to study the properties of this network. In particular, it is possible to enumerate and count the various types of rings within the structure of liquid water using a ring perception algorithm as described by Matsumoto et al. \cite{Matsumoto2007}. The algorithm functions as follows: (a) All triplets of connected vertices in the graph are stored in a list (b) For each triplet, a cyclic path search is performed, excluding shortcut paths. The search is completed when either a cyclic path is found, or the maximum ring size limit is reached (c) Repeat the same process for each triplet in the graph, and discard duplicate cycles. This allows one to obtain a distribution of the specific ring and fragment sizes resident in the system. To perform this analysis, the algorithm implemented in the \textit{cycless} Python package \cite{Matsumoto2007,cycless} was used to obtain the number of $K$-membered rings in the system, denoted by $N_K$, where $K=3,4,...,12$. This enumeration was done for every frame in the trajectory, resulting in a timeseries data of ring statistics. To normalize this data, the relative fraction of $K$-membered rings, $f_K = N_K/\sum_K N_K$ was calculated for each frame, and this timeseries data was used for further analysis. Description and raw data for all the ring-based analyses have been provided in the \textit{Supplementary Information (SI)}.

\subsection{Two-point time correlation function and memory kernel calculation}
The memory kernel formalism put forward by Meyer et al. \cite{Meyer2017,Meyer2021,Meyer2020} is the underlying principle behind the calculation of the memory kernel $K(\tprime,t)$. Briefly, the formalism can be explained as follows: consider a phase-space observable $A$, in this case, being the time-dependent order parameter. The nsGLE can be written as follows, providing the time-evolution of this observable \cite{Meyer2017,Meyer2019,Meyer2020,Meyer2021} (henceforth denoted simply by $A$):

\begin{align*}
    \frac{dA(t)}{dt} = \omega(t)A(t) + \int_0^t d\tprime K(\tprime,t) A(\tprime) + \eta(t)
\end{align*}

where $K(\tprime,t)$ is the memory kernel in consideration and $\omega(t),\eta(t)$ are functions that describe different observables for different processes. In general, the phase space averages $\langle A(t) \rangle$ of time dependent variables are done over the initial phase space distribution $\rho_0$ i.e. averaging over MD trajectories with different initial configurations and velocities corresponding to a fixed $\rho_0$. For the subsequent analysis, the two-time autocorrelation function of the observable $A$ is of interest, hence one calculates $C(\tprime,t) = \langle A^*(\tprime)A(t) \rangle$. By applying the nsGLE of $A(t)$, the equation of $C(\tprime,t)$ reads:

\begin{align*}
    \frac{\partial C(\tprime,t)}{\partial \tprime} = \omega(t)C(\tprime,t) + \int_{\tprime}^t d\tau C(\tprime,t) K(\tau,t)
\end{align*}

From the knowledge of $C(\tprime,t)$, the construction of $K(\tprime,t)$ is not straightforward, and requires an iterative inversion procedure. However, due to the discretized nature of the data in the time domain as obtained from MD simulations, it is possible to write it as a matrix system of equations, which allows for a numerical scheme that does not require iterative sums and can be handled by inversion of matrices. The numerical procedure put forward by \cite{Meyer2020} has been applied, outlined as follows. First, the two-point time correlation function (TCF) $C(\tprime,t)$ has been calculated for each individual trajectory by calculating the TCF of $f_6(t)$ with itself. For calculation of the TCF, the original timeseries $f_6(t)$ is put through a moving average filter with a window size of 2 ns. Having calculated the TCF for every trajectory, a statistical average is taken over all runs to arrive at an average $C(\tprime,t)$ which will be considered for further analysis. To normalize this TCF, the divisor is taken to be $\sqrt{C(t,t)C(\tprime,\tprime)}$, so that all diagonal elements are unity. For further analysis, as it will be required to calculate time derivatives through finite difference, the number of data-points in the TCF are increased to decrease the incidence of finite difference errors. This was done by using a bivariate cubic spline interpolation to make the grid finer by a factor of 100 (10 on each axis). The resulting $C(\tprime,t)$ for each $T_i$ has been shown in Figure S6 of the \textit{Supplementary Information} (SI). 

Once the $C(\tprime,t)$ is obtained, a matrix algorithm as described in \cite{Meyer2020} was used to calculate $K(\tprime,t)$. Due to the discrete nature of the time series data from MD trajectories, it is evaluated at integer steps of $\Delta t$, which is the grid size of the $C(\tprime,t)$ that has already been evaluated (in this case, 100 ps). Hence, the TCF has a matrix representation throughout this calculation, and in the following algorithm, the matrix representation will be used explicitly using the notation $\textbf{C}_{ij} = C(i\Delta t,j\Delta t)$. \textbf{Bold} symbols will be used to refer to matrices hereafter. One defines follows:

\begin{align*}
    \textbf{A} &= \text{diag}[C_{11}^{-1},C_{22}^{-1}...C_{NN}^{-1}] \\
    \textbf{B}_{ij} &= \textbf{C}_{ii} \\
    \textbf{S}_0 &= \textbf{ADC} \\
    \textbf{j}_0 &= \textbf{AD(B-C)}
\end{align*}

where $\textbf{D}_{ij} = (2\Delta t)^{-1}(\delta_{i+1,j} - \delta_{i-1,j})$ is the first finite difference matrix. At the boundaries of the matrix, forward and backward differences were considered, as applicable. Subsequently, one defines:

\begin{align*}
    \textbf{J}^U &= (\textbf{I} - \textbf{S}_0^U)^{-1} \textbf{j}_0^U \\
    \textbf{J}^L &= (\textbf{I} + \textbf{S}_0^L)^{-1} \textbf{j}_0^L \\
    \textbf{J} &= \textbf{J}^U + \textbf{J}^L
\end{align*}

where $U$ and $L$ denote the upper triangular and lower triangular parts, respectively. The convention chosen here is that only the upper triangular part will contain the diagonal to prevent double-summing the diagonal. Finally, one calculates:

\begin{align*}
    \textbf{K} = \textbf{J} \textbf{D}^T
\end{align*}

where $T$ denotes the matrix transpose. This $\textbf{K}$ matrix is equivalent to $K(\tprime,t)$ obtained at discrete grid points through the simple mapping $\textbf{K}_{ij} = K(i\Delta t,j\Delta t)$.

%%%%%%%%%%%%%%%%%%%%%%  FIGURE 1
\begin{figure*}[ht!]
\centering
\includegraphics[width=\textwidth]{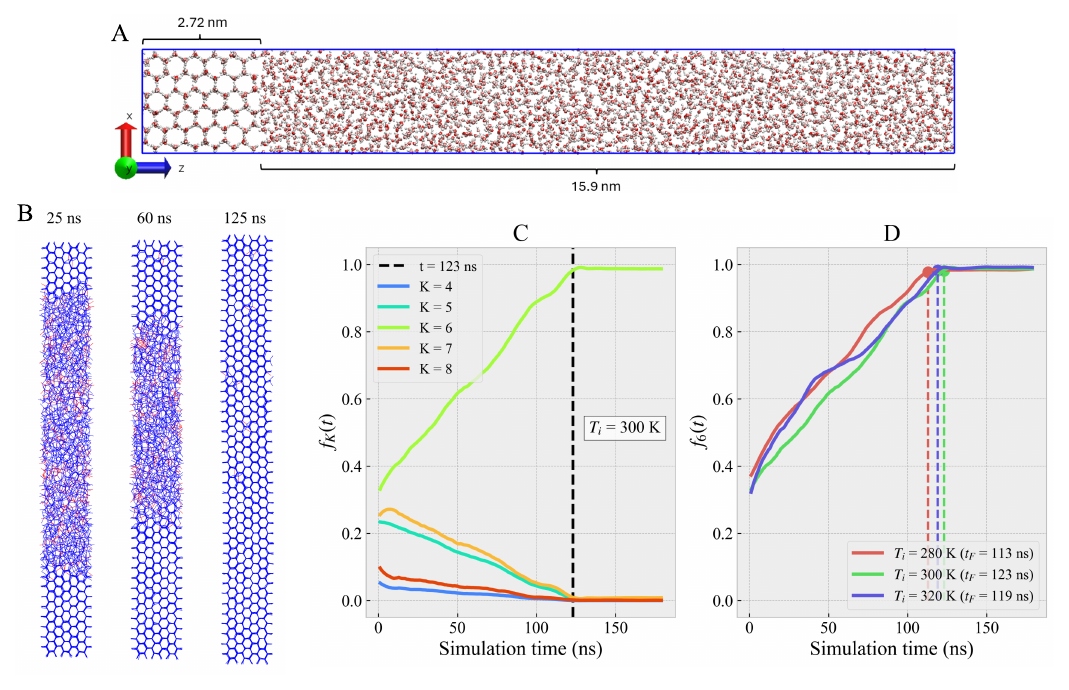}
\caption{Visualizations and analysis schematics. \textbf{(A)} The starting configuration for the direct coexistence simulations obtained by putting the ice configuration in contact with the liquid configuration, keeping the interface normal along the z-axis. The simulation box dimensions are 2.36 nm, 2.22 nm, 18.62 nm in the x, y and z directions respectively. \textbf{(B)} Visualizations of the system (obtained from one individual run with $T_i=300$ K) at various times during the production run. From left to right: 25ns, 60ns, 125ns. The \textcolor{blue}{blue} lines denote the hexagonal rings in the system, and the \textcolor{red}{red} lines denote pentagonal rings in the system. \textbf{(C)} Plot showing the time evolution of $f_K(t)$ for $K$ = 4, 5, 6, 7, 8 for one individual run with $T_i=300$ K. The vertical black dotted line has a horizontal intercept of 123 ns, the calculated $t_F$ for this trajectory. It can be seen that this line intersects the $f_K(t)$ curves at the point(s) where the ring distribution has converged to a delta distribution around $K=6$. \textbf{(D)} Plot showing the fraction $f_6$ of 6-membered rings as a function of simulation time for three different $T_i$: 280 K, 300 K, 320 K for one run. The freezing time $t_F$ is listed within parentheses inside the legend, and it's location on the plot is signified by a filled circle at the point where $f_6$ $\approx 1$. 
}
\label{fig:figure1}
\end{figure*}
%%%%%%%%%%%%%%%%%%%%%%  FIGURE 1

% SECTION HEADING
\section{Results}
\subsection{Freezing times versus initial temperature}
The system's evolution from a liquid/ice coexistence to a fully frozen state can be visualized over time by examining snapshots of the simulation trajectory at different time points along the trajectory (Figure \ref{fig:figure1}B). As the simulation progresses, the fraction of the hexagonal ice phase expands as water molecules freeze, ultimately resulting in complete solidification. 

To quantitatively track the freezing process, the fraction of 6-membered rings ($f_6$) was monitored as a function of simulation time. This quantity exhibited an increasing trend, indicating crystal growth and progressive freezing. After a sufficiently long time, $f_6$ stabilizes, fluctuating around a constant mean value close to unity, signifying that the system is fully frozen. The instant of time at which this stabilization occurs is defined as the ‘freezing time’ ($t_F$), which was compared across different initial liquid temperatures $T_i$. Through this procedure, the time series data of $f_K(t)$ was obtained for a given $T_i$.

Shown in Figure \ref{fig:figure1}C is the time series data $f_K(t)$ for one such run, with $T_i=300$ K. It can be seen that with the progression of simulation time, $f_6(t)$ increases towards unity, signifying the increasing number of 6-membered rings due to the growth of the hexagonal ice. Simultaneously, $f_{K\neq6}(t)$ decreases towards zero. Around the 123 ns mark, $f_6 \rightarrow 1$ and $f_{K\neq6} \rightarrow 0$. In other words, the ring size distribution $f_K$ has converged to a delta distribution centered at $K=6$. This implies that the only ring-type present in the system is the 6-membered ring, indicating that the system has become fully ice.

Simulations were conducted for the following $T_i$: 275 K, 280 K, 290 K, 300 K, 320 K, 340 K, and 360 K. For each $T_i$, 20 simulations were ran. From these runs, statistics of $t_F$ versus $T_i$ were collected, and further analyses were performed on these results. 

Shown in Figure \ref{fig:figure1}D is a plot generated for $f_6$ versus time for three values of $T_i$. It could be observed that $f_6$ increases with time, until it reaches unity when the system completely becomes ice. The instant of time where $f_6$ reaches its maximum value ($\approx 1$) is considered the $t_F$ of the system. For different starting temperatures, $t_F$ is expected to be different. For example, in Figure \ref{fig:figure1}D, the values of $t_F$ for $T_i=$ 280 K, 300 K, 320 K are 113 ns, 123 ns and 119 ns, respectively. In this way, multiple estimates of $t_F$ for the entire range of $T_i$ were obtained. The final curve showing the variation of $t_F$ as a function of $T_i$ was obtained by averaging over these estimates across 20 runs for each $T_i$.

%%%%%%%%%%%%%%%%%%%%%%  FIGURE 2
\begin{figure*}[ht!]
\centering
\includegraphics[width=\textwidth]{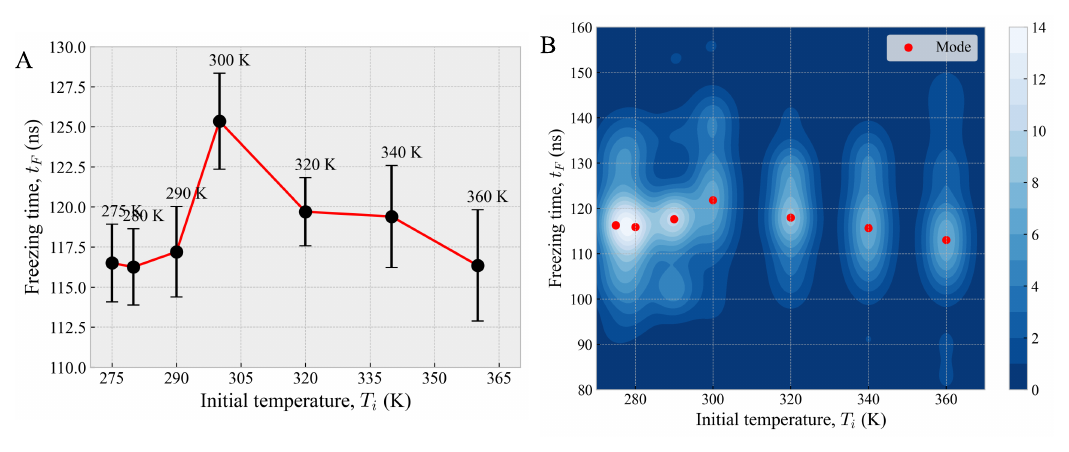}
\caption{
Freezing kinetics in the system via measurement of $t_F$ versus $T_i$. \textbf{(A)} Plot of the initial temperature $T_i$ (in Kelvin; horizontal axis) versus the freezing time $t_F$ (in nanoseconds; vertical axis). The \textbf{black} dots represent the mean value of $t_F$ for the corresponding $T_i$, The mean was obtained as the arithmetic mean from 20 runs. The error bars correspond to the standard error of the mean (SEM). \textbf{(B)} Heatmap plot of freezing times ($t_F$, in nanoseconds) on the x-axis and the starting temperature ($T_i$, in Kelvin) on the y-axis. Every individual datapoint $(t_F,T_i)$ is represented as a 2-D Gaussian of fixed spread, and the final heatmap is generated as the cumulative sum of all such Gaussians. The colourbar on the right represents the range of values within the heatmap. The red points represent the \textit{mode} of the observations for a fixed $T_i$ by locating the maxima of the heatmap at that particular $T_i$. 
}
\label{fig:figure2}
\end{figure*}
%%%%%%%%%%%%%%%%%%%%%%  FIGURE 2

Shown in Figure \ref{fig:figure2}A, is the plot of the mean freezing time $t_F$ versus initial temperature $T_i$. It can be observed that the freezing time $t_F$ exhibits a non-monotonous dependence on the initial temperature $T_i$. $t_F$ exhibits a low value at 275 K, increasing to a maximum value at 300 K, and decreasing with further increase in $T_i$ as shown by both the trend of the mean values. To observe the underlying statistics behind the non-monotonous dependence on $T_i$, a two-dimensional heatmap was generated by tallying $t_F$ versus $T_i$ for all 20 runs. For every individual datapoint $(t_F,T_i)$, a Gaussian of unit height and a fixed spread was added to the heatmap. The individual estimates of $t_F$ for all runs have been provided in the \textit{Supplementary Information (SI)}. The final heatmap was the total sum of every datapoint over all runs, as shown in Figure \ref{fig:figure2}B. Further, the most-occurring observation, or the mode was calculated for each $T_i$ by taking the local maxima of the heatmap at that specific $T_i$, and shown on the heatmap itself in Figure \ref{fig:figure2}B. It can be seen from the same that the mode observation follows the trend of the mean, showing a maximum at $T_i=300$ K, and decreasing on either side. Overall, the variation of the non-equilibrium freezing kinetics, measured by the freezing time, shows a non-monotonous behaviour with the initial temperature, as shown by both the average and mode of the statistics. The future of the system, as well as the time it takes to reach it, is dependent on the history, i.e. the initial state at which it starts of from; in this case it being the temperature of the liquid pre-quench. This indicates the presence of an inherent memory effect in the system via which the future evolution of the system is able to be influenced by the past state(s). 

\subsection{Memory kernel analysis}
The memory effect in the process can be quantified and compared by leveraging the so-called \textit{memory kernel} $K(\tprime,t)$ using the memory kernel formalism \cite{Meyer2017,Glatzel2021,Meyer2020,Meyer2021}. The role played by this two-point temporal function is via the memory term in the nsGLE, which describes the non-Markovian evolution of a phase-space observable via an integro-differential equation. This quantity has been feasibly used to define and quantify the memory effect of various processes \cite{Meyer2019,Meyer2020,Meyer2021}, including phase transitions as well. In general, the extent of support of the kernel in time is an indicator of the extent of temporary memory exhibited by the observable in question. By calculating the memory kernel for different cases, it becomes possible to examine the extent to which the memory of the system contributes to the time evolution of any phase-space observable of interest. In this work, the order parameter $f_6$ will be used as the relevant phase-space observable, and its evolution as a function of time across the various runs and initial temperatures will be examined under the lens of this formalism. 

Following the prescription in \cite{Meyer2020}, the two-point time correlation function $C(\tprime,t)$ of the order parameter $f_6(t)$ was calculated for each $T_i$ individually across the dataset. An average of $C(\tprime,t)$ was taken as the average over the dataset, and considered to be the final $C(\tprime,t)$ for the respective $T_i$. Afterwards, via a matrix algorithm as outlined in the \textit{Methods}, the memory kernel(s) $K(\tprime,t)$ were obtained for the range of $T_i$ in consideration. While the two-point function $K(\tprime,t)$ encodes the memory between times $t$ and $\tprime$, examining the entire function by plotting it across two independent axes would make a visual analysis complicated due to the coarse nature of the data. However, were one only interested in how the memory of the initial condition affects the system across time, then it would be sufficient to visualize a 'slice' of this function by setting one of the time indices to $t=0$, and investigate the time evolution of this reduced function with time to get a measure of the evolution of how much and how long the system 'remembers' the initial condition. Moreover, the cumulative integral of this function could also be analyzed as a function of time to get an understanding of the accumulated memory within the evolution of the system. 

\begin{figure*}[ht!]
\centering
\includegraphics[width=\textwidth]{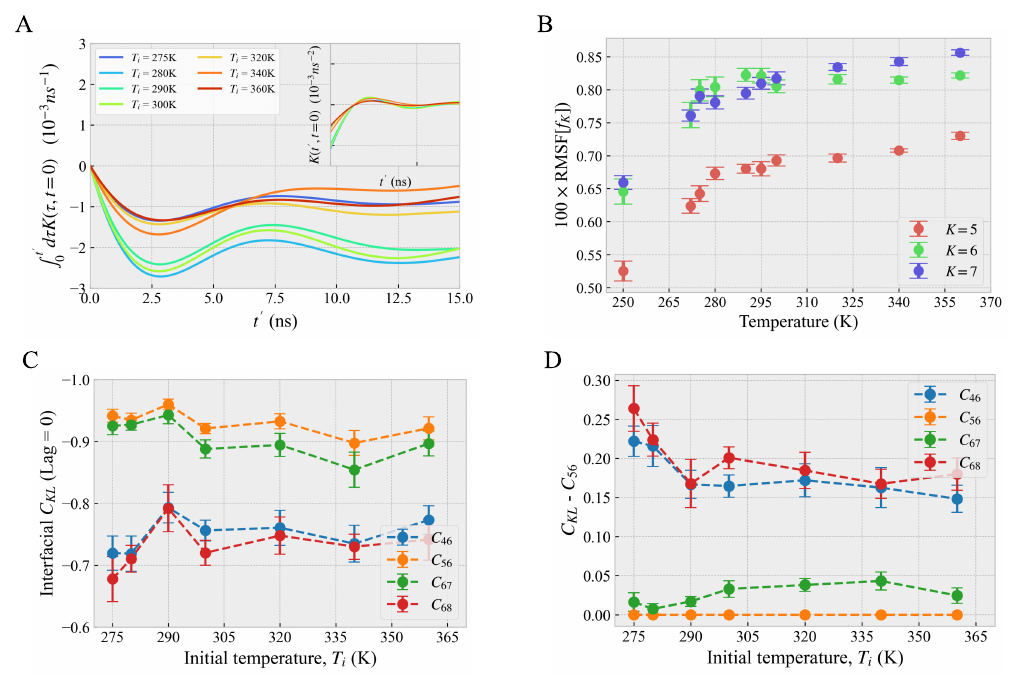} 
\caption{Temperature dependence of memory kernel and structural correlations at the ice-liquid interface. 
\textbf{(A)} Plot showing the cumulative integral of the memory kernel $\int_0^{\tprime} d\tprime K(\tprime,t=0)$ (vertical axis) against $\tprime$ (horizontal axis) for seven different values of $T_i$. 
\textbf{Inset:} Plot showing the variation of the memory kernel $K(\tprime,t=0)$ (vertical axis) against $\tprime$ (horizontal axis) for seven different values of $T_i$. The scales of both inset axes are the same as the external figure.
\textbf{(B)} Root mean squared fluctuations (RMSF) in $f_K$ obtained for various temperatures from equilibrium simulations. The error bars correspond to the standard error of the mean. 
\textbf{(C)} Interfacial ring cross-correlations quantified through $C_{KL}$, where K,L = $4,5,6,7,8$, and $C_{KL}$ is the zero-lag value of the cross-correlation of $f_K$ with $f_L$. Due to $f_6$ being anti-correlated with $f_{K\neq6}$, the cross-correlations are reported with a negative sign. The error bars correspond to the standard error of the mean. 
\textbf{(D)} Relative differences in the cross correlations $C_{6L} - C_{56}$, where L = $4,5,7,8$. The error bars correspond to the standard error of the mean.}
\label{fig:figure3}
\end{figure*}

As shown in Figure \ref{fig:figure3}A inset, the variation of $K(\tprime,t=0)$ with respect to $\tprime$ is plotted for several values of $T_i$. The curves indicate that the kernel function begins at a finite, nonzero value. With increasing $\tprime$, this value decays towards zero, suggesting that the system’s memory of the initial condition diminishes over time, albeit to different extents depending on $T_i$. This indicates a memory effect, which could be understood as a retardation in the growth of $f_6(t)$, i.e., the ice growth, consistent with the negative sign of the kernel. Notably, for $T_i = 280, 290$ and $300$ K, the kernel starts at a higher magnitude compared to the other cases. This manifests as persistence in the memory effect, seen in the integrated kernel. Figure \ref{fig:figure3}A illustrates this by showing the cumulative integral, $\int_0^{\tprime} d\tau K(\tau,t=0)$, as a function of $\tprime$. These curves highlight that the ‘accumulated’ memory is indeed more pronounced for $T_i = 280, 290,$ and $300$ K than for the other cases, reinforcing the observation that the memory contribution to the evolution of $f_6$ varies with the initial temperature.

From these observations, it seems that the memory kernel provides useful insights into the temperature-dependence of the freezing kinetics. For instance, in the case of $T_i=300$ K, the relatively long freezing time ($t_F$) coincides with the particularly strong memory retention reflected in the kernel results. However, the situation is not fully captured by this correspondence alone, as $T_i=280$ K and $290$ K also show significant memory retention, though with shorter $t_F$ compared to 300 K. It is important to note that the kernel analysis discussed so far only probes the dependence of the state at time $t$ on the state at $t=0$, leaving the influence of intermediate times $0 < \tprime < t$ less well resolved. Equally important is the nature of the initial condition itself: liquid water exhibits a non-monotonic temperature dependence of density, with a temperature of maximum density (TMD) that varies among models. For TIP4P/Ice, the TMD is $\approx 295$ K \cite{Vega2005}, close to the 300 K case. Since ice Ih has a lower density than liquid water, it is necessary that water undergoes a density change to form ice Ih. From a density point of view, the initial state at $T_i$ close to TMD is the farthest away from ice Ih. The combined interplay between the intrinsic memory effects quantified by the kernel and the features of the initial state together shape the overall freezing dynamics.

The non-monotonic variation of non-equilibrium freezing kinetics manifests as a maximum in the vicinity of the 300 K range and appears to reflect the trend originally observed by Mpemba \cite{mpemba1969}, who reported that freezing times show a maximum at $T_i = 20^{\circ}C$ (or 293 K), followed by a decrease at higher temperatures. While the exact location of the maximum temperature does not match precisely, there is a qualitative agreement with the statement of the effect itself. It is worth noting that later studies on the temperature trend of this effect place the location of the maximum differently to the original work, but fairly commensurate with the result here \cite{Ahn2016,Ghosh2025}. Combining the results of the freezing time statistics with the memory kernel analysis, it seems reasonable to conclude that the nonequilibrium process studied in this work displays characteristics consistent with the Mpemba effect. Importantly, the analysis suggests that memory effects play a key role in this phenomenon, and one might deduce that the Mpemba effect appears here as a consequence of the underlying memory dynamics. This interpretation aligns with existing literature where the Mpemba effect has been shown to be a memory-related phenomenon in various other systems \cite{Patron2023,Santos2024,Mompo2021,BaityJesi2019}, and the present work indicates that it may similarly manifest through the memory of nonequilibrium freezing processes in water. Notably, a recent study on the homogeneous nucleation of ice from supercooled water reveals non-Markovian effects in said process \cite{deHijes2026}, further lending interest to the central idea of memory effects.

In the aforementioned results, the memory kernel was extracted from the evolution of the $f_6$ order parameter, which is related to the fraction of hexagonal rings in the hydrogen bond network. While an overall coarse-grained picture of the system's evolution has been discussed, a deeper understanding of the molecular-level mechanisms underlying this behaviour is also desired. Simulation studies have made efforts to probe the microscopic source of this effect, and have proposed a range of mechanisms, ranging from the non-equipartition dissipation of thermal energy \cite{Gijon2019}, the populations of hexamer states in water \cite{Jin2015}, to the memory \cite{Zhang2014} or differing relative strengths of the hydrogen bond \cite{Tao2016} in water.. Given that the order parameter in focus here is fundamentally based on hydrogen bonding, it becomes natural to examine the network structure itself in greater detail. In the following section, further analyses of the bond network will be discussed to elucidate some of the underlying mechanics of the phenomena that have already been observed.

%% SECTION HEADING
\section{Analysis of hydrogen-bonded ring structures}
The prevalence and role of ring structures in liquid water was first examined within the Random Network Model (RNM) framework by Rice, Belch, and Sceats, establishing a basis for considering closed-loop hydrogen-bonded motifs as key microscopic variables in the liquid state \cite{Rice1981}. A theoretical argument was subsequently advanced by Speedy to attribute anomalies in heat capacity, compressibility, and thermal expansion to pentagonal rings, thereby elevating five-membered structures as central to water’s thermodynamic anomalies \cite{Speedy1984}. Confirmation through simulation was provided by Speedy and Mezei, who detected and characterized pentagonal rings and proposed their mutual stabilization and clustering, implying correlated organization in the network \cite{Speedy1985}. From this perspective, it was reasoned that liquid-to-ice transformation should be influenced by rearrangements that convert pentagons into hexagons, since hexagonal rings are structurally required for hexagonal ice \cite{Speedy1985}.

Interconversion among ring sizes was further analyzed by Belch and Rice using bond–bond correlations to infer time scales of correlated molecular motions in ring-bearing environments \cite{Belch1987}. The conversion between 6-membered and K-membered rings ($K\neq6$) was argued to be a fundamental process in liquid water, extending the Rice–Belch–Sceats conjecture and suggesting multiple interconversion mechanisms beyond a single dominant pathway \cite{Rice1981,Belch1987}. Within this framework, transformations that supply 6-membered rings are emphasized as central to crystallization kinetics, with particular attention placed on pathways that originate from pentagonal precursors \cite{Belch1987}.

The transformation of supercooled liquid at a crystal interface can be understood as governed by two influences: (a) the intrinsic ring-network dynamics in liquid water and (b) the external driving force provided by the ice interface. To interrogate the first aspect, a focus was placed on pentagonal rings due to their population being comparable to hexagons at the temperatures studied, unlike ring sizes farther from six that are less represented \cite{Belch1987, Formanek2020, Gao2021}. Population distributions across ring sizes indicate that decreasing temperature enhances $f_K$ for $K=5,6,7$ while reducing other $f_K$, sharpening the peak around $K=6$ in $P(\{f_K\})$, consistent with prior studies over temperature ranges \cite{Belch1987, Formanek2020, Gao2021}.

At $T=250$ K, the supercooled liquid is dominated by 5-, 6-, and 7-membered rings, while the crystal interface supplies the energetic driving force favouring conversion toward 6-membered rings at the advancing front. Consequently, the overall thermodynamics and kinetics of crystal growth are expected to be determined by the interconversion processes among these dominant ring sizes at the interface. It will be shown that pentagonal rings constitute the limiting factor for growth and that their behaviour underlies the observed correlation trends with respect to $T_i$ reported earlier in this work.

To understand the intrinsic dynamics of the liquid network, equilibrium MD simulations of pure water were performed to compute time-resolved ring statistics, to consider fluctuations that reflect ring stability. Root-mean-squared fluctuations, RMSF$[f_K]$, were calculated from trajectories recorded every 50 fs over 100 ps and averaged over 20 independent runs (details in Methods and SI), revealing consistently lower RMSF for $K=5$ relative to $K=6,7$ in Figure $\ref{fig:figure3}B$. This reduced fluctuation magnitude was interpreted as greater stability for pentagonal rings in the liquid, suggesting a kinetic impediment to their conversion into hexagons at the interface. This conclusion aligned with quantum chemical energy calculations \cite{Yang2019} at the MP2/6-311++G(d,p) level, which found the binding energy per molecule lower for $K=5$ than for $K=6$.

Having established the 5-membered ring as the limiting factor due to its enhanced stability, the interplay with the interface-driven bias toward hexagon formation was next investigated at $T=250$ K; where the final outcome is complete hexagonalization ($f_6\to1$) independent of kinetic pathway. Given that $f_6(t)$ must increase toward unity while $f_K(t)$ for $K \neq 6$ must decay to zero (as in Figure $\ref{fig:figure1}D$), anti-correlated temporal behaviour between hexagons and non-hexagons is implied by mass balance in ring space. This suggests a cross-correlation analysis between ring-fraction time series as a lens on the operative pathways during growth.

To quantify this, the direct coexistence system (Figure \ref{fig:figure1}A) was simulated sufficiently long for the interface region to crystallize (details in \textit{Methods} and \textit{Supplementary Information}). From this, the time evolution of $f_K(t)$ was obtained, and zero-lag cross-correlation coefficients $C_{KL}(0)$ for $K,L=4,5,6,7,8$ were calculated and averaged over 10 runs. Figure \ref{fig:figure3}C shows that cross-correlations between hexagon-pentagon and hexagon-heptagon rings were stronger at lower $T_i$ and decreased as $T_i$ increased. The (anti-)correlations $C_{56}$ and $C_{67}$ were greater compared to $C_{46}$ and $C_{68}$. This behaviour, together with the $K=5$ ring being limiting, explains why high $T_i$ corresponded to faster crystal growth rather than slower. Strong anti-correlation at lower $T_i$ indicated hexagonal rings mainly formed by interconversion with 5- or 7-membered rings. At higher $T_i$, the weaker anti-correlation and broader $f_K$ distribution implied that rings of types $K \neq 5,7$ also transformed to $K=6$, accelerating hexagonal crystal growth without any contribution from the $K=5$ intermediaries.

Furthermore, the differences between $C_{56}$ and $C_{46}$, as well as between $C_{56}$ and $C_{68}$, were found to decrease with $T_i$, indicating a growing relative contribution of $K=4,8$ rings to hexagon formation at higher interfacial temperatures (Figure $\ref{fig:figure3}D$). The elevated values of $C_{56}$ at lower $T_i$ point to the dominant influence of pentagonal stability on growth kinetics in colder interfaces, an influence that diminishes as $T_i$ rises and correlations weaken, allowing alternative pathways to participate more substantially, resulting in freer conversions to hexagons. The results reported in a temperature-quench study \cite{Jin2015} show that quenching from higher temperatures show higher retention of hexamer states as compared to quenching from lower temperatures, which indicates a similar behaviour to what one observes here. Taken in conjunction with each other, these correlation patterns clarify how ring-network stability and interfacial driving jointly determine pathway usage and thereby modulate crystal growth rates across $T_i$.

%%% SECTION 
\section{Conclusions}
In this study, simulations of the freezing process of water were carried out which elucidated the complex interplay between structural memory and non-equilibrium freezing kinetics. It has been demonstrated that the initial thermal history significantly influences freezing rates through imprinted microscopic structures in the hydrogen-bond network. Quantitatively, the freezing time $t_F$ exhibits a pronounced non-monotonic dependence on the initial temperature $T_i$, reflecting a memory effect where the initial configurations modulate the subsequent diffusion-limited ice growth dynamics.

The memory effect has been rigorously characterized through the non-stationary generalized Langevin equation framework by extracting two-time memory kernels $K(t', t)$ from the order parameter $f_6(t)$ measuring six-membered ring populations. The kernels begin at finite nonzero values with temporal evolution and integrated magnitudes that vary with $T_i$, indicating that the system "remembers" its initial structural state to various extents. This non-Markovian memory has been identified as a key factor in governing the kinetic retardation or acceleration of crystal growth, linking structural relaxation pathways with macroscopic freezing times.

A corollary to these findings is the qualitative reproduction of the Mpemba effect: the counterintuitive phenomenon where water initially heated to a higher temperature freezes faster than cooler water when quenched to the same final temperature. Although the precise temperature of maximum freezing time does not exactly match experimental reports, the observed non-monotonic freezing behavior and memory dynamics in simulations suggest that the Mpemba effect emerges naturally from the intrinsic memory encoded in water’s structural evolution. This reinforces the broader significance of memory effects in out-of-equilibrium phase transitions and encourages further experimental and theoretical exploration of thermal history effects in diverse complex systems.

At a molecular level, the hydrogen-bonded network ring structures have been analyzed, revealing that five-membered rings present a stable and kinetically persistent motif that acts as a bottleneck to crystallization. Their low root-mean-squared fluctuations and strong anti-correlations with six-membered rings at lower $T_i$ constrain the rearrangement processes necessary for ice formation. Conversely, at higher $T_i$, the weakening of these correlations enables alternative structural pathways that circumvent pentagonal ring stability, thereby facilitating faster ice growth. This mechanistic understanding connects the temporal evolution of network topology with the observed memory effects and freezing kinetics.

\section*{Supplementary Material}
A \textit{Supplementary Information} (SI) document containing the a description of the force field parameters, the ring analysis algorithm, determination of freezing time, and the calculation of cross-correlation coefficients has been provided. It also contains Tables and Figures containing all the raw data that contributed to the analysis presented here.

\section*{Acknowledgments}
The authors acknowledge IISc Bangalore STC (Space Technology Cell) for research grants and funding (Project code: ISTC/PHY/PKM/500). The authors thank BRNS (Board of Research in Nuclear Sciences) and SERB (Science and Engineering Research Board, now ANRF, Anusandhan National Research Foundation) for research grants and funding. The authors also thank the PKM-GPU2 HPC cluster at the Department of Physics, Indian Institute of Science for providing computing resources. AH and PKM thank R. Rajesh from the Institute of Mathematical Sciences (IMSc), Chennai and Subir K. Das from Jawaharlal Nehru Centre for Advanced Scientific Research (JNCASR) for key discussions and exchange of ideas during the course of this work.

\section*{Conflict of Interest Statement}
The authors have no conflicts to disclose.

\section*{Data Availability Statement}
The data that support the findings of this study are available within the article [and its supplementary material].

% The following command includes non-cited refs in the bibliography, commented out because we don't need it.
% \nocite{*}
\section*{References}
\bibliographystyle{plain}
\bibliography{refs} % Produces the bibliography via BibTeX.

\end{document}